\documentclass[superscriptaddress,twocolumn,floats,showpacs,prl,amsmath,amssymb,floatfix,nofootinbib,balancelastpage]{revtex4}

\input epsf
\usepackage{psfig}

\newcommand{\mnras}{Mon. Not. R. Astr. Soc.}

\begin{document}

\title{Modifications to the Cosmic 21-cm Background Frequency Spectrum by Scattering via electrons in Galaxy Clusters}
\author{Asantha Cooray}
\email{acooray@uci.edu}
\affiliation{Department of Physics and Astronomy, 4186 Frederick Reines Hall, University of California, Irvine, CA 92697}

\begin{abstract}

The cosmic 21-cm background frequency spectrum related to the spin-flip transition of neutral Hydrogen present  during and before the era of reionization
is rich in features associated with physical processes that govern transitions between the two spin states.
The intervening electrons in foreground galaxy clusters inversely Compton scatter the 21-cm background spectrum and modify it
just as the cosmic microwave background (CMB) spectrum is modified by inverse-Compton scattering. Towards typical galaxy clusters
at low redshifts, the resulting modification is a  few tenths milli-Kelvin correction to the few tens milli-Kelvin temperature of 21-cm
signal relative to that of the cosmic microwave background black body spectrum. The modifications are mostly associated with
sharp changes in the cosmic 21-cm background spectrum such as due to the onset of a Lyman-$\alpha$ radiation field or heating
of neutral gas. Though low frequency radio interferometers that are now planned for 21-cm anisotropy measurements are insensitive to
the mean 21-cm spectrum, differential observations of galaxy clusters with these interferometers can be utilized to indirectly establish
global features in the 21-cm frequency spectrum. We discuss the feasibility to detect the spectrum modified by clusters and find that for
upcoming interferometers, while a  detection towards an individual cluster is challenging, one can average signals over a number of clusters,
selected based on the strength of the Sunyave-Zel'dovich effect at high radio frequencies involving CMB scattering alone,
to establish the mean 21-cm  spectrum.


\end{abstract}


\maketitle

\noindent \emph{Introduction--- } The cosmic 21-cm background involving spin-flip line emission or absorption 
of neutral Hydrogen contains unique signatures 
on how the neutral gas evolved since last scattering at a redshift of 1100 and the subsequent reionization due to UV emission from stars and quasars at redshifts between 6 and 20 \cite{barkana}. 
The global signatures that are imprinted on the 21-cm background as a function of redshift are measurable today in terms of
the brightness temperature frequency spectrum relative to the black-body cosmic microwave background (CMB). 
Subsequent to recombination, the temperature of neutral gas is coupled to that of the CMB. At redshifts below
$\sim$ 200 the gas cools adiabatically, its temperature drops below that of the CMB, and neutral Hydrogen
resonantly absorbs CMB flux through the spin-flip transition \cite{field,loeb}. At much lower redshifts, gas temperature is
expected to heat up again as luminous sources turn on and their UV and soft X-ray photons reionize and heat
the gas \cite{chen}. A signature is also expected  from the Lyman-$\alpha$ radiation field produced by first sources \cite{barkana2} through the
Wouthuysen-Field effect \cite{Wouthuysen}.
While challenging, an observational measurement of  the mean 21-cm frequency spectrum, in the form of
modifications to the black body CMB spectrum, provides unique insights into the physics of cosmic 
neutral gas and the reionization history of the Universe \cite{shaver}.

The challenges for making a precise measurement of the spectrum come from 
the fact that it involves a total intensity measurement on the sky
rather than differential measurements that are easily pursued to study anisotropies in the intensity.
An accurate measurement of the brightness temperature spectrum requires a precise calibration of the instrument,
an accounting of galactic foregrounds and point sources,  and a control of systematics at the level of a few to a few tens mK \cite{shaver}.
Low frequency radio interferometers that are now pursued for 21-cm observations, such as
the Low Frequency Array (LOFAR; http://www.lofar.org), Miluera Wide-field Array (MWA; http://www.haystack.mit.edu/arrays/MWA),
and the Primeval Structure Telescope (PAST; \cite{Pen:05}), are not sensitive to the mean spectrum as
anisotropy observations inherently remove the mean signal on the sky. These interferometers directly measure
spatial inhomogeneities of the 21-cm intensity, which are induced by  fluctuations in the neutral Hydrogen density, temperature, and
 velocity, as well as spatial variations in the coupling between spin temperature and gas temperature
such as due to inhomogeneities in the Lyman-$\alpha$ radiation field. 
Given anisotropy observations, either in the form of the three-dimensional power spectrum of 21-cm signal or the two dimensional
angular power spectrum  when averaged over a redshift interval or a corresponding frequency bin,
detailed models of inhomogeneities are required to establish physical properties of reionization as well as the
mean 21-cm signal corresponding to that redshift or frequency bin \cite{zaldarriaga}.
There are large degeneracies between parameters that govern the mean brightness spectrum and those that lead to inhomogeneities \cite{cooray3}.
If signatures associated with the mean brightness temperature spectrum can be established independent of spatial variations,
then some of the degeneracies are broken. More importantly, however, the mean frequency spectrum alone can reveal when the
Universe ionized and on average when the Lyman-$\alpha$ radiation field became an important source of ionization
and when, if any, heating that happened to the gas.

In this paper, we discuss the possibility to use the modification to the cosmic 21-cm background frequency spectrum
by scattering via intervening electrons in galaxy clusters to 
indirectly establish global features in the mean 21-cm spectrum generated during and prior to reionization. The proposed observations can be
carried out with interferometers since the modification associated with low-redshift scattering can be
established from differential observations towards and away from galaxy clusters. Unlike an experiment to directly establish the cosmic 21-cm frequency
spectrum at low radio frequencies involving a total intensity measurement on the sky,
the differential observations with an  interferometer are less affected by issues such as the exact calibration 
of the observed intensity using an external source and the confusion from galactic foregrounds that are uniform over angular scales larger than a typical cluster.
The latter is the general case with the Galactic synchrotron background that is the signal at these low frequencies.
The resulting modification to the 21-cm spectrum towards a typical galaxy cluster at low redshifts is
at the level of a few tenths mK, while the original spectrum has 21-cm related signatures
with amplitudes at the level of a few tens mK relative to the blackbody CMB.
Such a small modification challenges an easy detection, but for upcoming interferometers, a detection could be achieved by averaging signals towards
a sample of galaxy clusters. Such multi-cluster observations are easily facilitated by the fact that the
instantaneous field-of-view of upcoming interferometers is expected to be more than 100 square
degrees and one expects in the order of hundred or more massive clusters in such fields.  
We begin our discussion with a summary of photon scattering via electrons in a cluster.

\begin{figure}[!t]
\centerline{\psfig{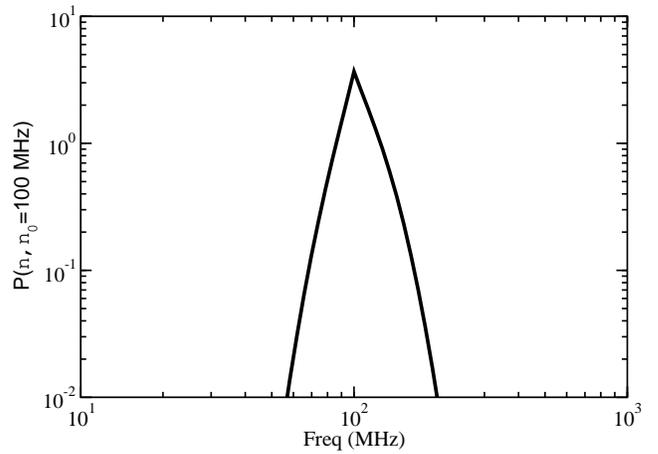}}
\caption{The scattering kernel $P_1(\nu,\nu_0)$ with $\nu_0=100$ MHz for a cluster with $k_B T_e=7$ KeV; 
For a different $\nu_0$, kernel can be shifted in $\log(\nu/\nu_0)$.  Any sharp features in the pre-scattered spectrum with variations
over a frequency range less than the typical width of the scattering kernel  will be broadened after scattering. The difference leads
to a decrement followed by an increment if the incoming radiation has a dip on the black-body CMB spectrum, while the opposite
happens if the pre-scattered spectrum has an increase on top of the CMB spectrum.}
\label{kernel}
\end{figure}

\noindent \emph{Scattering in Clusters--- } 
The inverse-Compton scattering of CMB photons via thermal electrons in galaxy clusters, generally referred to as the
Sunyaev-Zel'dovich effect (SZ; \cite{SunZel80}), is a powerful probe of cluster physics and cosmology
\cite{Birkinshaw,Carlstrom}. The changes to the CMB spectrum resulting from scattering  can be studied based on the Kompaneets equation
based on the  blackbody shape of the input spectrum \cite{Kompaneets}.
The  Sunyaev-Zel'dovich spectrum is that of a decrement and an increment, relative to CMB,  with  a 
cross-over at a frequency of 217 GHz; 
This difference in the spectrum relative to thermal CMB allows a variety of studies in multifrequency 
CMB anisotropy maps \cite{Cooray}. 

With modifications imposed to the CMB spectrum by neutral Hydrogen, an exact numerical calculation can be made to establish the
low frequency spectrum emerging from a cluster after scattering, $I(\nu)$, given an input spectrum $I_0(\nu)=2k_B \nu^2 T_b/c^2$ in the Rayleigh-Jeans (RJ) end of the frequency range. Since the input spectrum is no longer a black-body,
the standard approach for the CMB SZ description 
involving the Kompaneets equation, which leads to a frequency dependent term scales by
the Compton-y parameter, cannot be used here. 

Using the inverse-Compton scattering kernel,  $P_\tau(\nu,\nu_0)$, giving the probability 
that scattering has moved a photon from a frequency $\nu_0$ to $\nu$ when
the optical depth to scattering is $\tau$,
the scattered spectrum can be written as (see, Ref.~\cite{Birkinshaw} for details)
\begin{equation}
I(\nu) = \int_0^{\infty} d\nu_0\, P_\tau(\nu,\nu_0) I_0(\nu_0) \, .
\end{equation}
The scattering probability
can be calculated following Ref.~\cite{chan}. In the case of scattering via a single electron moving with speed 
$\beta c$, this
probability is
\begin{eqnarray}
&&P_1(\nu,\nu_0) = \frac{3}{16 \gamma^4 \beta} \\
&\times& \int_{\mu_{\rm l}}^{\mu_{\rm u}} d\mu \frac{(1+\beta \mu_1)}{(1-\beta \mu)^3}
\left[1+\mu^2 \mu_1^2 +\frac{1}{2} (1-\mu^2)(1-\mu_1^2)\right] \, , \nonumber
\label{eqn:psingle}
\end{eqnarray}
where $\gamma = (1-\beta^2)^{-1/2}$, $\mu_1 = [\nu(1-\beta \mu) -\nu_0)]/\beta \nu_0$, the lower limit of the integral $\mu_{\rm l}$
is either $-1$ when $\nu \leq \nu_0$ or $(\nu - \nu_0(1+\beta))/\beta \nu$ otherwise, and the upper limit of the integral
$\mu_{\rm u}$ is either $1$ when $\nu \geq \nu_0$ or $(\nu - \nu_0(1-\beta))/\beta \nu$ otherwise.

For scattering via a  population of electrons, as in the case of galaxy clusters,
the scattering probability in equation~(2) must be averaged over the probability distribution of electron velocities,
$P_e(\beta)$:
\begin{equation}
P(\nu,\nu_0) = \int_{\rm \beta_{\rm lim}}^1 d\beta\; P_e(\beta) P_1(\nu,\nu_0) \, ,
\end{equation}
where $\beta_{\rm lim}$ is $|\nu - \nu_0|/(\nu+\nu_0)$. We assume a thermal electron population in clusters
with an electron temperature $T_e$  such that the velocity
distribution is given by
\begin{equation}
P_e(\beta) = \frac{\gamma^5 \beta^2 m_e c^2 \exp(-\gamma m_ec^2/k_B T_e)}{k_B T_e K_2(m_e c^2/k_B T_e)} \, ,
\end{equation}
where $K_n(x)$ is the modified Bessel function of the second kind. The scattering kernel $P(\nu,\nu_0)$
is illustrated in Figure~1.
While we have considered the single scattering case, to allow for multiple scattering given the optical depth.
If $\tau << 1$, we can write
the scattering probability as $P_\tau(\nu,\nu_0) = (1-\tau)\delta(\nu - \nu_0) + \tau P(\nu,\nu_0)$ \cite{Birkinshaw,Taylor}. 
In the case of CMB, the numerical evaluation of Eqs. (1) to (4) has been  extended to study both
relativistic corrections \cite{Rephaeli} and the variation associated with scattering by non-thermal electrons in clusters \cite{Blasi};
For the present discussion, we ignore such complications as these only lead to minor corrections on top of the standard calculation.

\begin{figure}[!t]
\centerline{\psfig{file=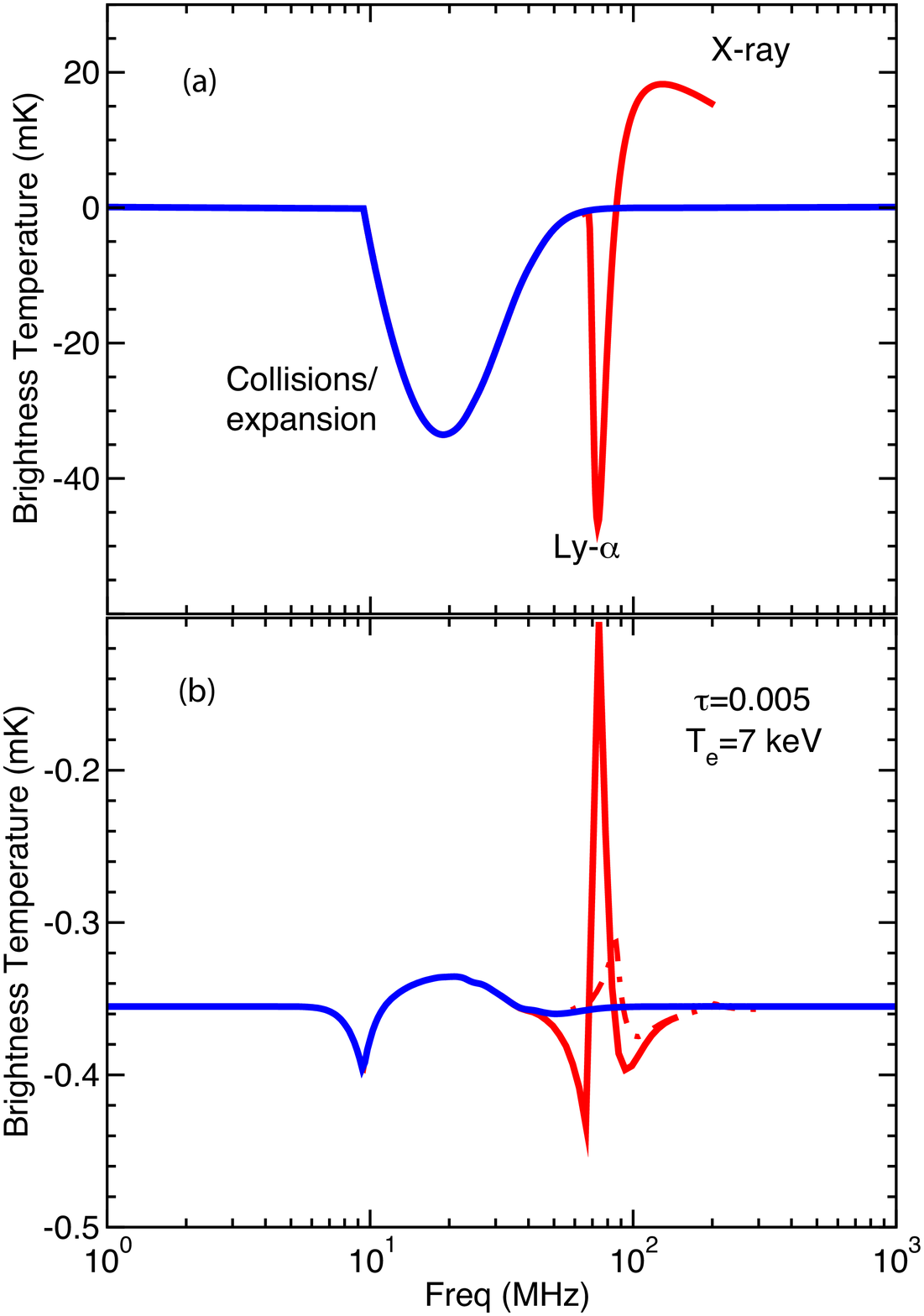,width=3.3in,angle=0}}
\caption{(a) The incoming radiation spectrum plotted in terms of the brightness temperature  relative to CMB.  
Three features are shown: A broad absorption due to 
collisional coupling  between spin and gas temperatures when  $30 < z < 200$\cite{loeb},
coupling  induced by  Ly-$\alpha$ radiation from first luminous sources 
when $20 < z < 30$ \cite{barkana2} (labeled Ly-$\alpha$), and an increase in the gas temperature when heated by soft X-rays
during reionization at a redshift below 20 \cite{chen} (labeled X-ray). 
(b) The difference spectrum after scattering through a galaxy cluster
with an electron temperature of 7 keV and an optical depth of 0.005, consistent with 
known clusters at low redshifts \cite{Mason} with the optical depth profile smeared over a beam of 5 arcminutes to be consistent with typical
beam sizes expected for upcoming interferometers.  The overall offset in the spectrum, at a brightness 
temperature $\sim$ -0.35 mK with no frequency dependence at the Rayleigh-Jeans tail, 
is due to scattering of the CMB black-body with no 21-cm signatures.
The 21-cm signatures lead to variations  up to 0.2 mK,
While the broad dip at very low frequencies leads to a decrement and an increment,
 Ly-$\alpha$ and X-ray features combine to given an overall increase towards clusters
at a frequency around $\sim$ 75 MHz; This is associated with the sharp dip related to the Lyman-$\alpha$
induced coupling of gas and spin temperatures at a redshift around 20. The dot-dashed line is the
scattered spectrum only involving the broad dip and the X-ray heating increment; the latter leads to an increment followed
by a decrement.}
\label{spectra}
\end{figure}

\begin{figure}[t]
\centerline{\psfig{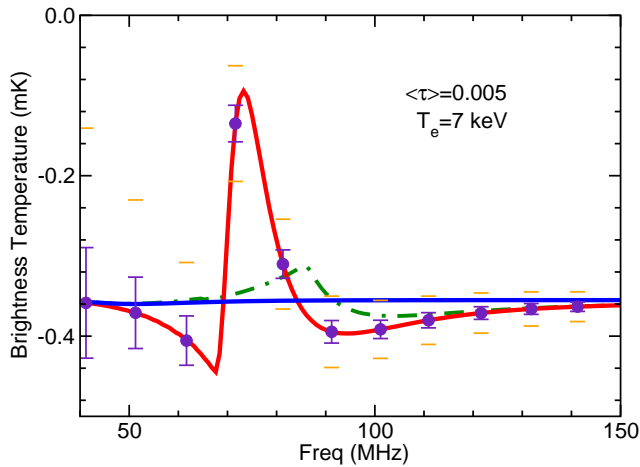}}
\caption{The brightness temperature difference spectrum towards and away from a cluster with a beam-smeared
average optical depth of 0.005 and an electron temperature  of 7 keV
that can be targeted with upcoming low-frequency radio interferometers (same as Figure~2b);
the optical depth and electron temperature values are typical of low redshift clusters \cite{Mason}.
Since underlying 21-cm signatures are global, 
one can also stack spectra towards a large sample of clusters to  improve the signal-to-noise ratio.
The errors bars assume experimental parameters similar to SKA with a system noise temperature of 1000 K 
at 100 MHz, and a scaling of $\nu^{-2}$ to other frequencies, a 5 MHz  bandwidth, synthesized beam of 5$'$,
and a year-long integration. The larger errors indicated by horizontal lines assume observations towards
an individual cluster, while error bars show the improvement by averaging over
observations of 10 clusters, though in planned observations $\sim$ few hundred massive clusters will be imaged in a given field-of-view
instantaneously. A large uncertainty in establishing the exact signal-to-noise for detection is the
foreground distribution within clusters, such as point sources, and the extent to which
such sources can be removed \cite{Morales}. }
\label{error}
\end{figure}

\noindent \emph{Results--- }  In Figure~2, we summarize the input and the output spectra of the 21-cm radiation towards a cluster
in terms of the brightness temperature that measures the difference with respect to CMB.
Current descriptions of the 21-cm radiation suggest three potentially interesting signatures.
These involve a broad dip as neutral Hydrogen resonantly absorbs CMB intensity through spin-flip transition 
at a redshift range between 200 and 30 \cite{loeb}, a feature related to the Lyman-$\alpha$ radiation field produced by first sources 
at redshifts between 30 and 20 \cite{barkana2}, and during reionization (at redshifts between 6 and 20),
an increment in the brightness temperature as soft X-ray photons from stars and quasars begin to heat the gas \cite{chen}.
Since the 21-cm background is formed by a line emission or an absorption, features that are imposed as a function of redshift map to
a unique frequency in the brightness temperature spectrum observable with telescopes today. 
The brightness temperature variations produced by above modifications to CMB by neutral Hydrogen
are generally at the level of 20 mK to 40 mK.

Inhomogeneities in the neutral Hydrogen distribution, its temperature, and the intensity of the Lyman-$\alpha$ photon radiation field
that couples spin temperature to the gas temperature via Wouthuysen-Field effect \cite{Wouthuysen}
generate anisotropies in the brightness temperature during and prior to reionization.
These fluctuations are at the level of a few mK to tens of mK \cite{zaldarriaga}. A measurement of these anisotropies,
in the form of an angular power spectrum is now considered to be one
of the main goals of planned low-frequency interferometers. Additional fluctuations are also generated during the
propagation such as due to gravitational lensing \cite{cooray2} and as we discuss later through scattering in galaxy clusters.
The intrinsic anisotropies, however, are dominated by those related to
low radio frequency foregrounds and techniques are now developed to remove confusing sources \cite{Morales}.

The brightness temperature difference towards and away from a cluster is shown in Figure~2(b).
First, note the overall amplitude difference between the two panels.
While the brightness temperature in the unscattered spectrum is generally at a level of
30 mK, the brightness temperature spectrum after scattering is
 an overall shift of about -0.35 ($\tau$/0.005) mK, independent of the frequency,  with
variations of order 0.05 mK to 0.2 mK imprinted on it. As an example here, we selected a cluster with an optical
depth $\tau$ of 0.005 and an electron temperature of 7 keV. These values are
 consistent with known electron distribution of low-redshift clusters \cite{Mason} when the optical depth profile is
averaged over a beam size of 5 arcminutes. The 5 arcminutes beam size is the typical of what is expected for upcoming
low frequency radio interferometers at a frequency of 100 MHz \cite{Morales}.

The overall offset in the mean temperature to a value of $-0.35$ mK is the usual SZ effect associated with the
black-body CMB spectrum at the Rayleigh-Jeans part of the frequency spectrum.
On top of this signal one finds 21-cm signatures in the form of decrements and increments.
The broad absorption in the redshift range between 200 and 30 modifies to a
sharp decrement of $0.05(\tau/0.005)$ mK followed by an increment of $<$ $0.05(\tau/0.005)$ mK.
While such very low frequencies will not be observable with upcoming radio interferometers, 
in the frequency range easily accessible to first generation 21-cm interferometers is the
signature related to reionization. Since the gas heating increment quickly follows the Ly-$\alpha$ related dip,
the overall change after scattering in clusters is a decrement followed by a sharp increment and another decrement. 
The latter signatures are model dependent as they require assumptions on when sources turn on,
the intensity of the Lyman-$\alpha$  radiation produced by these sources, 
and the intensity of the soft X-ray background that heats the gas. We also considered the case where Ly-$\alpha$ signature is
not present, but only a signature related to gas heating (dot-dashed line in Figure~2b). In this case
modifications to the spectrum have smaller amplitudes, but is an increment followed by a decrement.
Clearly, the amplitudes and the cross-over frequencies are sensitive to exact details of the reionization process
and a measurement of the spectrum in Figure~2(b) should be  a goal for upcoming experiments.

If the pre-scattering brightness temperature spectrum has signatures at the level of 30 mK to 40 mK
over a narrow range in frequency, as in the Ly-$\alpha$ signature of Figure~2(a),
resulting signatures in the post-scattering brightness temperature spectrum
towards clusters are potentially measurable with current and upcoming interferometric 
arrays. To illustrate this possibility, in Figure~3, we plot the expected errors on the brightness temperature
assuming an array with a system temperature, $T_{\rm sys}$, of 1000 K at 100 MHz, 
with $T_{\rm sys} \propto \nu^{-2}$, a bandwidth $\Delta \nu$ of 10 MHz
for observations, an integration time of 1 year,
and a collecting area that of the Square Kilo-meter array (SKA) with
$A \sim 1 km^2$. For interferometric observations, the pixel noise in the synthesized beam is
\begin{eqnarray}
&&\Delta T = 0.03\; {\rm mK} \left(\frac{A}{A_{\rm SKA}}\right)^{-1} \left(\frac{\Delta \nu}{10 \; MHz}\right)^{-1/2} \\
&&\left(\frac{t_{\rm obs}}{1\; yr}\right)^{-1/2} \left(\frac{T_{\rm sys}}{1000\; K}\right) \left(\frac{\nu}{100\; MHz}\right)^{-2} 
\left(\frac{\Delta \theta}{5'}\right)^{-2} \, . \nonumber
\end{eqnarray}
Since the background spectrum  is the quantity of interest, given prior knowledge on the optical depth and 
electron temperature, one can appropriately scale and combine observations towards multiple clusters
to increase the signal-to-noise ratio. In this case, we assume the improvement in $\Delta T$ with $\sqrt{N_{\rm cl}}$
of the number of clusters $N_{\rm cl}$ used.

The errors shown in Figure~3 are those using a single cluster and the improvement by averaging over signals towards 10 clusters.
When integrated over the whole spectrum, relative to the case where the scattering effect is due to CMB alone,
one detects variations at the level of $\sim$ 2 $\sigma$ towards a single cluster, but better than 10 $\sigma$ when
averaged over $\sim$ 10 clusters. For smaller arrays that will be soon be operational,
such as MWA with an area 10 times smaller than the SKA, due to the large instantaneous fields-of-view of
 300 deg.$^2$ \cite{Morales}, $\sim$ 10$^3$ clusters will be imaged and a better than 5 $\sigma$ detection of
variations related to Ly-$\alpha$ coupling or X-ray heating can be obtained by stacking signals towards $\sim$ 100 clusters.
The stacking procedure requires prior knowledge on the cluster optical depth profile and the electron temperature.
This can be obtained a priori based on SZ observations related to the CMB spectrum alone at high radio frequencies.
In the near future, all-sky maps from Planck  will provide a substantial catalog of massive SZ clusters
with profiles averaged over the Planck's angular scale of 5 arcminutes \cite{planck}. Since the angular scale
naturally corresponds to those of low-frequency observations, one can use information on the high frequency SZ profile as a way to normalize
low frequency signals when stacking to establish signatures related to the 21-cm spectrum. 

The calculation related to the modification resulting from scattering in clusters assume that the 
incoming spectrum through the cluster is
same as the 21-cm frequency spectrum outside the cluster. Any difference between the two will lead to an  additional signal in the difference towards and away from the cluster. 
While such an extra difference can aid the detection, it can also 
complicate an easy interpretation of
the measurements especially when attempting to establish global signatures in the mean 21-cm spectrum as one 
must account for arcminute-scale anisotropy in the incoming signal itself. 
By carefully averaging signals over a large number of clusters, spatial variations in the incoming
signal may be accounted for. This scattering related differences towards and away from clusters
may also be used as a probe
of the typical size scale of reionization, such as due to inhomogeneities in the Lyman-$\alpha$ radiation field
from first UV sources. 

Studies towards a large sample of clusters is possible since
upcoming low-frequency interferometers will have instantaneous fields of view of order 300 deg.$^2$ or more.
One would naturally expect of order a few hundred clusters in such fields with adequate mass to produce
measurable differences. The same clusters will also be detectable at high radio frequencies with CMB experiments
such as Planck. The latter observations are useful to find suitable cluster candidates for low-frequency studies 
based on the strength of the SZ signal at high frequencies and an estimate of the optical depth profile, in combination
with X-ray estimates of the temperature. The SZ spectrum alone, at high-frequencies, may provide independent estimates of the temperature (at the level of a few keV), which may be useful for low-frequency cluster studies in the absence of X-ray data \cite{Hansen}.

Before concluding, we note that for all these studies, the main worry is  complications resulting from foregrounds.
Since observations are differential any smooth foreground at  angular scales more than the
cluster will be removed.  This will remove a significant fraction of the Galactic foreground signal.
The main worry is point 
sources within clusters and a potential smooth synchrotron background associated with electrons in clusters that
appear as radio halos and radio relics \cite{GioFer00}.
The foregrounds, however, are expected to be controlled through a combination of direct point-source removal, when resolved, and through spectral information; Techniques  that are developed to remove foregrounds in
anisotropy data can be easily applied for cluster observations \cite{Morales}.
Unfortunately, estimates on the level of confusion within clusters are highly uncertain
due to the lack of cluster observations at low radio frequencies.  
Upcoming arrays will provide a wealth of data to further understand the expected level of confusion. 

To summarize, 
the cosmic 21-cm background frequency spectrum related to the spin-flip transition of neutral Hydrogen present  during and before the era of reionization
is rich in features associated with physical processes that govern transitions between the two spin states.
The intervening electrons in foreground galaxy clusters inversely Compton scatter the 21-cm background spectrum and modify it
just as the cosmic microwave background (CMB) spectrum is modified by inverse-Compton scattering. 
While low-frequency interferometer arrays  are built or planned to primarily study 21-cm anisotropies, they can easily target
galaxy clusters to establish signatures in the 21-cm background. We strongly encourage an analysis of signals towards and away from clusters
as a way to establish signatures related to Lyman-$\alpha$ radiation intensity field or gas heating during and prior to reionization.
The analysis is challenging as the expected signal difference due to scattering is at most 0.2 mK for low redshift clusters with typical
optical depths of order 0.005 and may be further complicated  by the underlying anisotropies associated with topology of reionization itself.
The availability of imaging data for a large numbers of clusters ($\sim$ 10$^3$) instantaneously with low frequency radio
interferometers for 21-cm studies as well as the availability of SZ data for such clusters soon with experiments like Planck will
allow joint studies to be conducted.

\smallskip
\noindent \emph{Acknowledgments--- } Author thanks J. Pritchard for electronic tables of the spectrum shown in Figure~2(a)
and Chris Carilli, Miguel Morales, and Peter Shaver for comments.


\begin{thebibliography}{99}

\bibitem{barkana}
  R.~Barkana and A.~Loeb,
  Phys.\ Rept.\  {\bf 349}, 125 (2001).

\bibitem{field}
G.~B.~Field, \apj\ {\bf 129}, 536 (1959);
D.~Scott and M.~J.~Rees, \mnras\ {\bf 247}, 510 (1990)

\bibitem{loeb}
  A.~Loeb and M.~Zaldarriaga,
  Phys.\ Rev.\ Lett.\  {\bf 92}, 211301 (2004);
  S.~Bharadwaj and S.~S.~Ali,
  Mon.\ Not.\ Roy.\ Astron.\ Soc.\  {\bf 352}, 142 (2004).



\bibitem{chen}
X.~L.~Chen and J.~Miralda-Escude,
  Astrophys.\ J.\  {\bf 602}, 1 (2004).

\bibitem{barkana2}
R.~Barkana and A.~Loeb,
  Astrophys.\ J.\  {\bf 626}, 1 (2005).


\bibitem{Wouthuysen}
S.~A.~Wouthuysen, Astron. J.\ {\bf 57}, 31 (1952);
G.~B.~Field, Astrop. J.\ {\bf 129}, 525 (1959).





\bibitem{shaver}
  P.~A.~Shaver, R.~A.~Windhorst, P.~Madau and A.~G.~de Bruyn,
  Astron.\ Astrophys. {\bf 345}, 380 (1999).


\bibitem{LOFAR}
LOFAR: http://www.lofar.org; SKA: http://www.skatelescope.org.


\bibitem{Pen:05}
J.~B.~Peterson, U.~L.~Pen and X.~P.~Wu,
  arXiv:astro-ph/0502029.



\bibitem{zaldarriaga}
  M.~Zaldarriaga, S.~R.~Furlanetto and L.~Hernquist,
  Astrophys.\ J.\  {\bf 608}, 622 (2004);
  I.~T.~Iliev, E.~Scannapieco, H.~Martel and P.~R.~Shapiro,
  Mon.\ Not.\ Roy.\ Astron.\ Soc.\  {\bf 341}, 81 (2003).


\bibitem{cooray3}
 M.~McQuinn, O.~Zahn, M.~Zaldarriaga, L.~Hernquist and S.~R.~Furlanetto,
  arXiv:astro-ph/0512263;
  M.~Santos and A.~Cooray, in preparation (2006)

\bibitem{SunZel80}
R.~A.~Sunyaev and Ya.~B.~Zel'dovich, Mon.\ Not.\ Roy.\ Astron.\ Soc.\ {\bf 190},
 413 (1980).

\bibitem{Birkinshaw}
M.~Birkinshaw, Phys. Rept.\ {\bf 310}, 97 (1999).

\bibitem{Carlstrom}
  J.~E.~Carlstrom, G.~P.~Holder and E.~D.~Reese,
  Ann.\ Rev.\ Astron.\ Astrophys.\  {\bf 40}, 643 (2002).

\bibitem{Kompaneets}
A.~S.~Kompaneets, Sov. Phys. JETP, {\bf 4}, 730 (1957).

\bibitem{Cooray}
  A.~Cooray, W.~Hu and M.~Tegmark,
  Astrophys.\ J.\  {\bf 540}, 1 (2000).

\bibitem{chan} S. Chandrasekhar,  {\it Radiative Transfer} (New York: Dover) (1960);
  E.~L.~Wright, \apj\ {\bf 232}, 348 (1979).

\bibitem{Taylor}
G.~B.~Taylor \& E.~L.~Wright, \apj\ {\bf 339}, 619 (1989).

\bibitem{Rephaeli}
Y.~Rephaeli, Ann. Rev. Astro. \& Astrop. {\bf 33}, 541 (1995).

\bibitem{Blasi}
 P.~Blasi, A.~V.~Olinto and A.~Stebbins,
  arXiv:astro-ph/0001471.





\bibitem{cooray2}
  K.~Sigurdson and A.~Cooray,
  arXiv:astro-ph/0502549;
 A.~R.~Cooray,
  New Astron.\  {\bf 9}, 173 (2004).



\bibitem{Morales}
  M.~F.~Morales, J.~D.~Bowman and J.~N.~Hewitt,
  arXiv:astro-ph/0510027;
 M.~G.~Santos, A.~Cooray and L.~Knox,
  Astrophys.\ J.\  {\bf 625}, 575 (2005).


\bibitem{Mason}
 B.~S.~Mason and S.~T.~Myers,
  arXiv:astro-ph/9910438.


\bibitem{planck}
  J.~Geisbuesch, R.~Kneissl and M.~Hobson,
  Mon.\ Not.\ Roy.\ Astron.\ Soc.\  {\bf 360}, 41 (2005)
  [arXiv:astro-ph/0406190];
 B.~Malte Schafer and M.~Bartelmann,
  arXiv:astro-ph/0602406.

\bibitem{Hansen}
  S.~H.~Hansen, S.~Pastor and D.~V.~Semikoz,
  Astrophys.\ J.\  {\bf 573}, L69 (2002)
  [arXiv:astro-ph/0205295].

\bibitem{GioFer00}
    G.~Giovannini and L.~Feretti,
  arXiv:astro-ph/0008342.

\end{thebibliography}
\end{document}